\documentclass{aa}  

\usepackage{graphicx}
\usepackage{txfonts}
\usepackage{soul}
\usepackage{hyperref}
\hypersetup{
    colorlinks   = true, 
    urlcolor     = cyan, 
    linkcolor    = blue, 
    citecolor   = blue 
}
\usepackage{threeparttable}
\usepackage{placeins}
\usepackage{xcolor}
\usepackage{soul} 
\setstcolor{red} 
\newcommand{\msun}{M$_\odot$\,}
\newcommand{\kms}{km s$^{-1}$}
\newcommand{\gyr}{$13~{\rm Gyr}~$}
\newcommand{\trh}{t_{\rm rh}}
\newcommand{\rc}{R_{\rm c}}
\newcommand{\rhl}{R_{\rm hl}}
\newcommand{\rlim}{R_{\rm lim}}
\newcommand{\rcrhl}{\rc/\rhl}
\newcommand{\mbh}{M_{\rm BH}}
\newcommand{\mgc}{M_{\rm GC}}
\newcommand{\vto}{V_{\rm TO}}
\newcommand{\mto}{m_{\rm TO}}

\newcommand{\qfit}{\texttt{QFIT}\,}
\newcommand{\deltaobsnew}[1]{\Delta_{\rm obs}(< #1)}
\newcommand{\deltaobs}{\Delta_{\rm obs}(< \rhl)}
\newcommand{\deltaobsadditional}{\Delta_{\rm obs}(<0.7 \rhl)}
\newcommand{\vdispratio}{\sigma_\mu\,(<0.2\rhl) / \sigma_\mu(\rhl)}
\newcommand{\equipart}{\mu(<0.1\rhl)}

\newcommand{\ngc}[1]{NGC~#1}
\newcommand{\figref}[1]{Fig.~\ref{#1}}
\newcommand{\myeqref}[1]{Eq.~\eqref{#1}}
\newcommand{\tabref}[1]{Table~\ref{#1}}
\newcommand{\secref}[1]{Sect.~\ref{#1}}

\begin{document}
\title{Inference of black-hole mass fraction in Galactic globular clusters}
\subtitle{A multi-dimensional approach to break the initial-condition degeneracies}
\titlerunning{Inferring the black-hole population in Galactic globular clusters}

\author{
    A. Della Croce \inst{1,2}\thanks{\email{alessandro.dellacroce@inaf.it}}
    \and 
    F. I. Aros\inst{3}
    \and 
    E. Vesperini\inst{3}
    \and
    E. Dalessandro\inst{2}
    \and
    B. Lanzoni\inst{1}
    \and
    F. R. Ferraro\inst{1}
    \and 
    B. Bhat\inst{1,2}
}

\institute{
    Department of Physics and Astronomy ‘Augusto Righi’, University of Bologna, via Gobetti 93/2, I-40129 Bologna, Italy
    \and
    INAF – Astrophysics and Space Science Observatory of Bologna, via Gobetti 93/3, I-40129 Bologna, Italy
    \and 
    Department of Astronomy, Indiana University, Swain West, 727 E. 3rd Street, IN 47405 Bloomington, USA
}

\date{Received \dots; accepted \dots}
 
\abstract
{Globular clusters (GCs) are suggested to host many stellar-mass black holes (BHs) at their centers, thus resulting in ideal testbeds for BH formation and retention theories.  
BHs are expected to play a major role in GC structural and dynamical evolution and their study has attracted a lot of attention. 
In recent years, several works attempted to constrain the BH mass fraction in GCs typically by comparing a single observable (for example mass segregation proxies) with scaling relations obtained from numerical simulations.}
{We aim to uncover the possible intrinsic degeneracies in determining the BH mass fraction from single dynamical parameters and identify the possible parameter combinations that are able to break these degeneracies.}
{We used a set of 101 Monte Carlo simulations sampling a large grid of initial conditions. 
In particular, we explored the impact of different BH natal kick prescriptions on widely adopted scaling relations. 
We then compared the results of our simulations with observations obtained using state-of-the-art HST photometric and astrometric catalogs for a sample of 30 Galactic GCs.
}
{
We find that using a single observable to infer the present-day BH mass fraction in GCs is degenerate, 
as similar values could be attained by simulations including different BH mass fractions.  
We argue that the combination of mass-segregation indicators with GC velocity dispersion ratios could help us to break this degeneracy efficiently. 
We show that such a combination of parameters can be derived with currently available data. However, the limited sample of stars with accurate kinematic measures and its impact on the overall errors
do not allow us to discern fully different scenarios yet.
}
{}

\keywords{
  methods: numerical
- stars: black holes
- stars: kinematics and dynamics
– globular clusters: general
- black hole physics
}
\maketitle

\section{Introduction\label{sec:intro}}
Massive stellar clusters are ideal cosmic laboratories of multi-body dynamics \citep{heggie_hut2003} and gravitational wave astrophysics \citep{rodriguez_etal2015,hong_etal2018}.
Due to asymmetric supernova explosion, BHs are expected to experience natal kicks \citep{janka_2013,mandel_2016}: if the kick amplitude is larger than the local escape speed, the BH is ejected.
The increasing number of detections of stellar mass BHs in Galactic GCs \citep[][see also \citealt{panuzzo_etal2024,balbinot_etal2024} for the recent detection of a 33~\msun BH in the ED~2 stellar stream]{maccarrone_etal2007,strader_etal2012,chomiuk_etal2013,miller-jones_etal2015,Giesers_etal2018} has thus triggered an extensive effort into the study of the long-term retention of BHs in massive clusters. 
These studies have shown that a sizeable population of BHs could be retained at the center of GCs for timescales longer than the Hubble time \citep{morscher_etal2013,sippel_hurley2013,heggie_giersz2014,arca-sedda_etal2018,askar_etal2018}. 
Also, 
following the recent discoveries of BH-star binaries by \emph{Gaia} \citep{tanikawa_etal2023,Chakrabarti_etal2023_bh1,el-badry_etal2023_bh2,el-badry_etal2023_bh1,panuzzo_etal2024}
young star clusters were investigated as possible BH formation grounds \citep{rastello_etal2023,tanikawa_etal2024a,tanikawa_etal2024,dicarlo_etal2024}. Within this context, BH retention in lower-mass star clusters would place strong constraints on BH natal kicks \citep[see e.g.,][]{torniamenti_etal2023}.

Several works addressed the role and impact of a population of BHs in the long-term dynamical evolution of a stellar system. They showed that the presence of BHs, specifically the heating from dynamically formed binary BHs, can significantly delay the mass segregation of visible stars and the core collapse of GCs \citep[see e.g.,][]{mackey_etal2007,mackey_etal2008,breen_heggie2013,Morscher_etal2015,alessandrini_etal2016,peuten_etal2016,weatherford_etal2018}.

However, the BH retention in GCs and the distribution of kick velocities after their formation are still matters of investigation 
\citep{Belczynski_etal2002,repetto_etal2012,janka_2013,mandel_2016,repetto_etal2017,giacobbo_mapelli_2020,andrews_kalogera_2022}.
The search for stellar-mass BHs in Galactic GCs, therefore, opens up a window on many fundamental and timely science cases, including the constraint of the early BH retention and natal kicks, the study of stellar dynamical interactions, 
up to the BH-BH merging in dense stellar systems as a source of gravitational wave emission 
\citep[][\citealt{arcsedda_etal2023_dragonII,arcasedda_etal2024_dragonI,arcasedda_etal2024_dragonIII,marinPina_etal2024,elBadry_etal2024}]{moody_sigurdsson2009,Banerjee_etal2010,rodriguez_etal2015,rodriguez_etal2016,rodriguez_etal2016b,rodriguez_etal2018,antonini_etal2016,hurley_etal2016,askar_etal2017,fragione_kocsis2018,hong_etal2018,samsing_dorazio2018,samsing_etal2018,zevin_etal2019}.

Recently, many studies addressed the inference of the total mass in stellar mass BHs harbored by GCs 
\citep{askar_etal2018,zocchi_etal2019,askar_askar_etal2019,weatherford_etal2020,dickson_etal2024}.
In particular, \citet{askar_etal2018} explored several correlations \citep[obtained from Monte Carlo simulations, see][]{arca-sedda_etal2018} to infer the properties of the BH subsystem, if present, using as observational anchor the luminosity density within the half-mass radius. The authors shortlisted 29 GCs that could potentially harbor a significant number (up to a few hundred) of BHs.
\citet{weatherford_etal2020} used a theoretical correlation between the fraction of BHs and the degree of mass segregation \citep{weatherford_etal2018} to infer the present-day BH population and their total mass in 50 Galactic GCs.
Finally, \citet{dickson_etal2024} performed multi-mass modeling of several cluster observables (i.e., velocity dispersion profiles along the proper motion (PM) and line-of-sight (LOS) directions, number density profile, and mass function measurements) for 34 GCs, thereby being able to constrain the total dynamical mass in dark remnants at the cluster centers. 
Interestingly, they found typically lower BH mass fractions compared to
\citet{askar_etal2018}, and \citet[][see e.g. Fig.~3 in \citealt{dickson_etal2024}]{weatherford_etal2020}.

As discussed by \citet[][see their Sect.~2.6]{askar_etal2018}, the inference of the mass in BHs using a single observable 
can be strongly biased and dependent on the specific assumptions adopted in the analysis.
In this work, we thus address the degeneracies in the inference of present-day BH population in GCs, possibly arising from multiple assumptions about the underlying physical processes that are still poorly constrained observationally.
The purpose of this work is therefore 1) to point out that some observable structural quantities used in the literature to infer the mass fraction in BHs are consistent both with systems with a significant mass fraction in BHs and systems with no (or a negligible fraction of) BHs; 2) to possibly identify the combination of dynamical parameters that allows unambiguous identification of the presence of a significant population of BHs. 

The paper is organized as follows: in \secref{sec:simulation} we present the simulation survey and its set of initial conditions, and in \secref{sec:results_simulation} we discuss the parameters investigated and the implications for BH mass fraction-inference in real GCs. In \secref{sec:observations} we describe 
a detailed comparison between the observations and our set of simulations, trying to disentangle different scenarios. In \secref{sec:conclusion} we summarize the results and conclude.
 
\section{Methods and initial conditions\label{sec:simulation}}
In this work, we use a set of 101 Monte Carlo simulations \citep{henon_etal1971a,henon_etal1971b} performed with the MOCCA\footnote{The name MOCCA stands for MOnte Carlo Cluster simulAtor, see \url{https://moccacode.net/}.} code 
\citep[][]{giersz_etal2013,hypki_etal2013}.
The MOCCA code follows the evolution of star clusters including the effects of two-body relaxation, single and binary stellar evolution 
\citep[modeled with the SSE and BSE prescriptions; ][]{hurley_etal2000,hurley_etal2002}, 
close stellar interactions \citep[which were integrated by using the FEWBODY code][]{fregeau_etal2004}, 
and a spatial cut-off modeling the effect of the tidal truncation due to the host galaxy.

The set of simulations analyzed in this work was fully presented in \citet{bhat_etal2024} in the context of defining novel structural parameters to 
determine the stage reached by GCs in their
evolution toward the core-collapse and post-core-collapse phases.
Here we briefly summarize the initial conditions adopted and refer to that paper for further details.
Each simulated cluster starts with an equilibrium configuration defined by a \citet{king_1966} distribution function, assuming a central dimensionless potential of $W_0 = 5$ or $7$. 
The truncation radii of our models are equal to the tidal radii of clusters moving on circular orbits in a logarithmic potential for the Galaxy at galactocentric distances of 2, 4, and 6 kpc.
A filling factor (defined as the ratio between the three-dimensional half-mass, $r_{\rm hm}$, and the tidal, $r_{\rm t}$, radii) of $0.025, 0.050$ or $0.1$ was adopted. 
The number of particles, $N_{\rm p}$ (defined as the sum of the number of single stars and binaries), varies between 500k, 750k, and one million with a $10\%$ primordial binary fraction. 
The initial distribution of binary properties was set following the eigenevolution procedure described in \citet{kroupa_1995} and \citet{kroupa_etal2013}.
Finally, for each set of these initial conditions, two simulations were performed assuming a different prescription for the BH natal kicks: 
either a Maxwellian distribution with a dispersion of 265~\kms \citep[i.e., the same as neutron stars, NSs, ][]{hobbs_etal2005}, or a reduced kick velocity based on the fallback prescription by \citet{Belczynski_etal2002}.
We adopted a 
\citet{kroupa_2001} stellar initial mass function between 0.1 and 100 \msun and a metallicity $Z=10^{-3}$.
Finally, we did not include 7 simulations in which an intermediate-mass black hole was formed. Studying the impact of intermediate-mass black hole formation on observable cluster properties is beyond the scope of this work, and will be the subject of future studies.

\section{Results\label{sec:results_simulation}}
\begin{figure*}[!th]
    \centering
    \includegraphics[width=\textwidth]{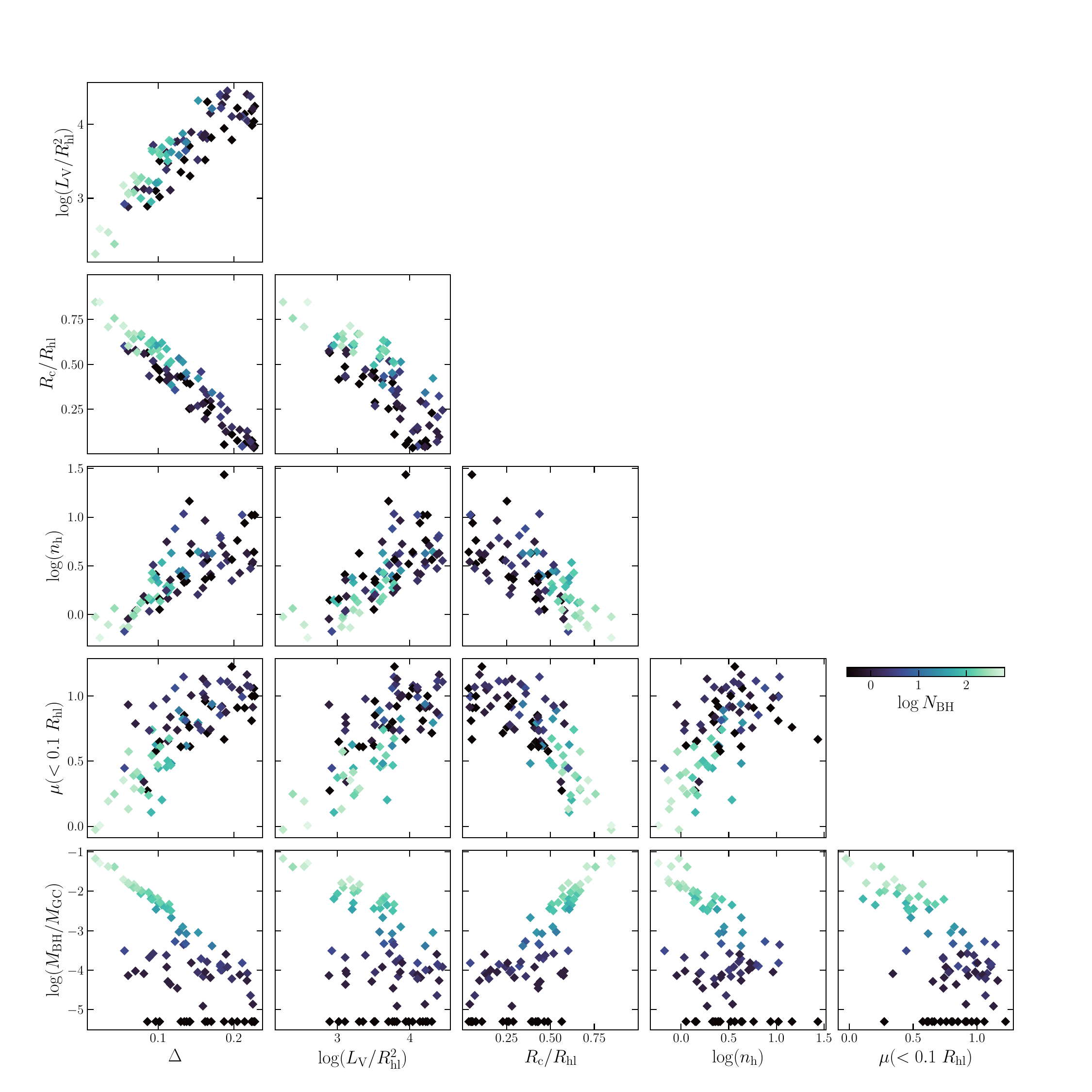}
    \caption{Simulation properties at 13~Gyr, each point representing a different simulation. Moving downwards on the $y-$axis: luminosity density (in units of $L_\odot~{\rm pc}^{-2}$), core to half-light radius ratio ($\rcrhl$), dynamical age ($n_{\rm h}$), inverse of the equipartition mass, $\equipart$, BH mass fraction ($\mbh/\mgc$), and mass segregation parameter $\Delta$. 
    Simulations are color-coded according to the number of BHs ($N_{\rm BH}$) at \gyr.
    A value of $\log N_{\rm BH} = -0.5$ was assigned to those with no BHs. 
    Similarly, simulations that do not retain any BH at \gyr are shown at $\mbh/\mgc = 5\times10^{-6}$ on the bottom row for visualization purposes only.
    }
    \label{fig:multipanel}
\end{figure*}

\begin{figure}[!th]
    \centering
    \includegraphics[width=0.47\textwidth]{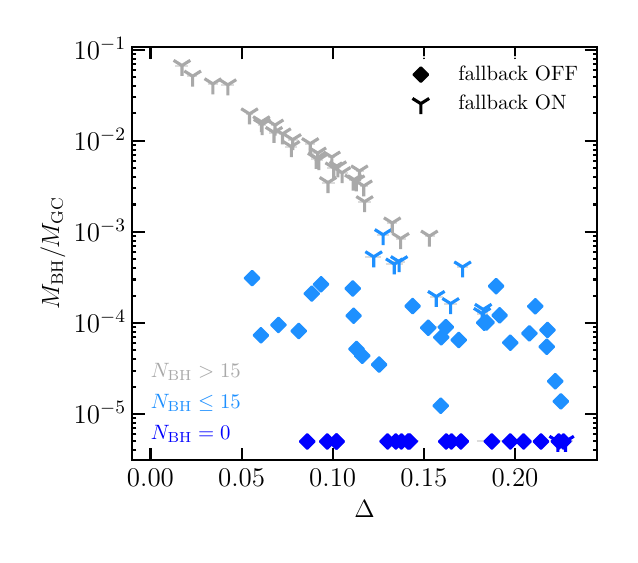}
    \caption{
    BH mass fraction as a function of $\Delta$. Colors depict the number of BHs in the simulation at \gyr, while different symbols show the adopted fallback prescription in the initial conditions. Simulations without BHs at \gyr are shown at $\mbh/\mgc = 5\times10^{-6}$. To account for fluctuations in the estimates arising due to the discrete nature of the simulations, we performed 500 different projections along the LOS, and for each of them, we measured $\Delta$. The values shown are the median ones. Errors obtained computing the 16\% and 84\% percentiles of the distribution are also shown although barely visible.
    }
    \label{fig:BHs_on_segregation}
\end{figure}



\begin{figure}[!th]
    \centering
    \includegraphics[width=0.47\textwidth]{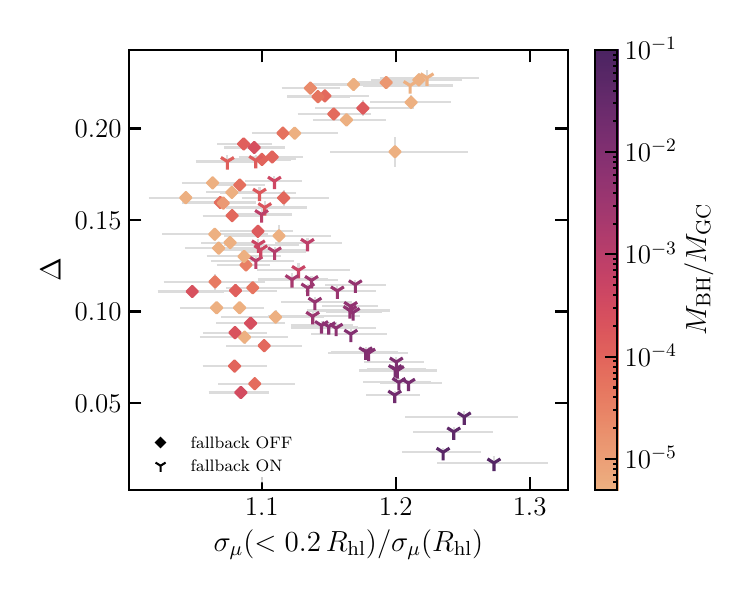}
    \caption{Mass segregation parameter as a function of the velocity dispersion ratio. Symbols show whether the simulation had (diamond) or had not (upside-down triangle) the fallback prescription for BH formation. 
    Different colors depict the BH mass fraction (see the color bar on the side)
    Finally, error bars were obtained from multiple projections along the LOS.}
    \label{fig:delta_veldispratio}
\end{figure}

In this Section, we present the analysis and the properties at \gyr of the simulations in the survey.
Firstly, for each simulation, we computed the total stellar mass ($\mgc$) and the BH mass fraction (defined as $\mbh/\mgc$, with $\mbh$ being the total mass in BHs).
We then explored the impact of long-lived BH subsystems on the evolution of star clusters by computing several quantities, some of which were previously used to infer the presence of massive BH populations in Galactic GCs \citep[see e.g.,][]{mackey_etal2008,bianchini_etal2016,askar_etal2018,weatherford_etal2020}: the mass-segregation parameter ($\Delta$), the average luminosity density, ($L_{\rm V} / \rhl^2$, with $\rhl$ the half-light radius and $L_{\rm V}$ the total V luminosity within $\rhl$),
the concentration ratio ($\rcrhl$ with $\rc$ being the core radius), the dynamical age ($n_{\rm h}$), and the inverse of the equipartition mass ($\mu$) estimated within $0.1\rhl$.

The parameter $\Delta$ was defined as
\begin{equation}
    \Delta = \int^1_0 {\rm nCRD}_{\rm pop1}(x) - {\rm nCRD}_{\rm pop2}(x)\,dx \,,
    \label{eq:delta_definition}
\end{equation}
where nCRD is the normalized cumulative radial distribution computed for a more massive (labeled as pop1) and a lower-mass (labeled as pop2) population.
The integral is computed between the cluster center and a limiting distance $\rlim = 2\rhl$ with $x$ corresponding to the projected distance of each star from the center, normalized to $\rlim$.
Similar parameters proved powerful tools in studying the properties and dynamical evolution of GCs \citep[see e.g.,][]{alessandrini_etal2016,lanzoni_etal2016,peuten_etal2016,ferraro_etal2018,ferraro_etal2019,ferraro_etal2023,ferraro_etal2023b,raso_etal2017,dalessandro_etal2019}, and were already used in previous studies to investigate the role of BH subpopulations within GCs \citep[see e.g.,][]{weatherford_etal2018}.
We exploited almost the full stellar mass range available at \gyr to maximize the mass-segregation signal: for the more massive population, we selected stars within $[\mto-0.025;\,\mto]$\,\msun (with $\mto = 0.8155$\,\msun being the main-sequence turn-off mass at \gyr for a simple stellar population with $Z=10^{-3}$), whereas 
for pop2 we selected stars in the mass range $[0.1;\,0.125]$~\msun.
According to the definition in \myeqref{eq:delta_definition}, $\Delta$ is a dimensionless parameter that traces the relative spatial concentration of massive stars compared to lower-mass ones through their nCRDs: the larger the value, the more massive stars are spatially segregated compared to the lower-mass ones.

The proxy for the dynamical age, $n_{\rm h}$, was defined as the ratio between the cluster physical age and the half-mass relaxation time \citep[$\trh$; ][]{spitzer_1987} calculated at that physical age; following \citet{spitzer_1987}:
\begin{equation}
    \trh = 0.138 \frac{\mgc^{1/2}\,r^{1/2}_{\rm hm}}{\langle m \rangle\,G^{1/2}\,\ln{\Lambda}}\,,
    \label{eq:half-mass_relaxation_time_spitzer1987}
\end{equation}
with $G$ the gravitational constant, $\langle m \rangle$ the mean star mass,
and $\Lambda$ the Coulomb logarithm coefficient. We used $\Lambda=0.02~N_{\rm p}$, which accounts for the effects of a mass spectrum \citep{giersz_heggie1996}.

To compute $\rc$, we fitted an analytical, multi-power law model to the surface brightness profile in the V band obtained using stars with $V<\vto+2$, with $\vto$ being the turn-off magnitude.
For each simulation, $\vto$ was estimated by dividing stars into magnitude bins (between $V=17-23$ mag, 0.1 mag wide) and selecting the bin with the bluest $V-I$ color. Finally, the core radius was obtained as the radius at which the surface brightness is half the central one. 
The half-light radius ($\rhl$) was defined as the radius enclosing half of the total projected light in the V band and computed directly from the simulation.

Finally, we computed the $\equipart$ parameter as presented in \citet{aros_vesperini2023}. This quantity represents the inverse of the equipartition mass \citep{bianchini_etal2016} with the advantage of providing a simpler description of the stellar-mass-dependence of the velocity dispersion \citep[see][for further details]{aros_vesperini2023}.

In \figref{fig:multipanel} we present all the possible combinations of the aforementioned parameters computed for each simulation.
In particular, the bottom row in \figref{fig:multipanel} shows the BH mass fraction as a function of different properties: the presence of a BH subsystem inhabiting the cluster central regions prevents the core collapse of visible stars thereby delaying the evolution of their structural properties 
\citep{mackey_etal2007,mackey_etal2008,breen_heggie2013,Morscher_etal2015}. 
This is in turn reflected in the $\rcrhl$ ratio \citep[which increases for higher BH mass fractions, see e.g.,][]{kremer_etal2020}, and the luminosity density \citep[which decreases for increasing BH mass fractions, see the discussion in][]{arca-sedda_etal2018}.

In \figref{fig:BHs_on_segregation} we focus on the BH mass fraction as a function of $\Delta$.
When simulations with the fallback prescription are considered, it is possible to observe a nice correlation between the BH mass fraction and $\Delta$, as expected based on the well-established role of binary BHs in halting the mass segregation of massive visible stars \citep[]{breen_heggie2013}. A similar trend was also recovered by \citet{weatherford_etal2018}.
However, \figref{fig:BHs_on_segregation} also shows that clusters can exhibit little mass segregation (i.e., low values of $\Delta$) without a sizeable population of BHs or even with no BHs at all. 
These systems have long initial relaxation times and BHs were ejected right after formation due to large natal kicks (for these simulations the fallback off prescription was adopted).
Hence, while mass segregation can provide us with valuable information, the role of different initial conditions and physical assumptions should be carefully considered
\citep[see also, for instance, the discussion in Sect. 2.6 of ][]{askar_etal2018}.
Also, a similar behavior is observed in all the quantities presented in \figref{fig:multipanel}.
The fact that clusters with very different BH mass fractions might exhibit similar properties (in terms of mass segregation, concentration ratio, and luminosity density see Fig.~\ref{fig:multipanel})  highlights the possible problems in inferring the BH mass fraction in real GCs using a single observable. Therefore, multiple physical properties should be jointly used to constrain the presence of BHs within GCs.

In this respect, we introduce a new observable parameter which in synergy with $\Delta$ turns out to be particularly useful in discriminating between BH retention and dynamical evolution. This is defined as $\vdispratio$, which is the ratio between the 1D velocity dispersion\footnote{
Following what is commonly done in PM studies, we used the two velocity components projected on the plane of the sky to determine the radial ($\sigma_{\rm R}$) and tangential ($\sigma_{\rm T}$) velocity dispersions for each cluster. We then defined $ \sigma^2_\mu \equiv (\sigma^2_{\rm R} + \sigma^2_{\rm T})/2$.} computed within $0.2\rhl$ and at $\rhl$\footnote{Computed from stars with projected distance from the center $\in[0.95; 1.05]~\rhl$.}.
This parameter quantifies the steepness of the velocity dispersion profile, which 
directly reflects the radial variation of the gravitational potential.

In \figref{fig:delta_veldispratio} we show the $\Delta-\vdispratio$ plot. Low mass segregation levels (roughly $\Delta<0.15$) can be explained either by the presence of a massive BH subsystem (darker points in \figref{fig:delta_veldispratio}) or due to the system being dynamically younger, without requiring high BH mass fractions (lighter points in \figref{fig:delta_veldispratio}, but see also \figref{fig:multipanel}). 
However, these two classes depart in $\vdispratio$: the presence of a massive BH subsystem deepens the potential well in the central regions, increasing the velocity dispersion ratio. 
On the other hand, dynamically young systems without many BHs exhibit lower values.

Velocity dispersion ratios on the order of $1.15$ are attained by systems with low BH mass fractions only if they are dynamically evolved (i.e., roughly $\Delta>0.2$), effectively breaking the degeneracy between dynamically young GCs without large BH mass fractions and systems hosting a massive BH population which slowed down their dynamical aging.

\section{Observations\label{sec:observations}}
In this Section, we present the observational analyses carried out for a sample of Galactic GCs to compare their structural and kinematical properties with those from simulations.

\subsection{Properties of Galactic GCs \label{sec:galactic_gc_properties}}
We selected Galactic GCs with both photometric data from \citet{sarajedini_etal2007} and individual-stars PMs by \citet{libralato_etal2022}\footnote{publicy available at \url{https://archive.stsci.edu/hlsp/hacks}}, covering at least the central $0.7 \rhl$. 
Such a selection included GCs proposed as promising candidates for hosting high BH mass fractions (including \ngc{5053}, \ngc{6101}, and \ngc{6362}, see \citealt{askar_etal2018,weatherford_etal2020}) while allowing to investigate a sufficiently large radial range. 
A large radial range enables better characterization of the system properties, such as mass segregation and the velocity dispersion profile.
Finally, we empirically found that selecting stars down to three magnitudes below $\vto$ was the best compromise between keeping the mass gap between pop1 and pop2 as large as possible and ensuring photometric completeness of at least $0.5$ over the whole radial range for a sizable fraction of GCs.
In \secref{appendix:completeness} we present the calculation of the photometric completeness and we discuss those cases for which the incompleteness was too severe and the calculation of $\Delta$-like quantities is not feasible with the dataset adopted in this work.
Out of the 57 clusters in common between \citet{sarajedini_etal2007} and \citet{libralato_etal2022}, 30 met all the aforementioned criteria.
For cluster ages, masses, and characteristic radii, we used the catalog provided by 
\citet[][but see also \citealt{vasiliev_baumgardt2021,baumgardt_vasiliev2021}]{baumgardt_etal2020}\footnote{The catalog is publicly available at \url{https://people.smp.uq.edu.au/HolgerBaumgardt/globular/}. Values used in this work are from the 4th version of the catalog updated in March 2023.}.

To estimate $\vdispratio$, we used the recent astrometric catalog by \citet{libralato_etal2022}.
The catalog consists of PMs and multi-epoch photometry for stars in about the central 100" of 56 Galactic GCs. We applied the same quality selections presented in Sect. 4 of \citet{libralato_etal2022} retaining stars with $V < \vto+1.25$ (as done for the simulations).
Adopting the same spatial selections (\secref{sec:results_simulation}), and using $\rhl$ from the catalog by \citet[][4th version]{baumgardt_etal2020}, we computed the 1D velocity dispersion accounting for errors on the $n$ individual stars. In particular, we assumed the likelihood function 
\citep{pryor_meylan1993}
\begin{align}
    \ln \mathcal{L} = \sum^{n}_{i=1}\, & -\frac{1}{2}\left( \frac{(v_{i,{\rm R}}-\langle v_{\rm R} \rangle)^2}{\sigma^2_{\rm R} + \delta v^2_{i,{\rm R}}} + \ln\,(\sigma^2_{\rm R} + \delta v^2_{i,{\rm R}})\right) +\nonumber \\
    & -\frac{1}{2}\left( \frac{(v_{i,{\rm T}}-\langle v_{\rm T} \rangle)^2}{\sigma^2_{\rm T} + \delta v^2_{i,{\rm T}}} + \ln\,(\sigma^2_{\rm T} + \delta v^2_{i,{\rm T}})\right) \,,
    \label{eq:likelihood}
\end{align}
with $v_{i,{\rm R}}, v_{i,{\rm T}}$ being the radial and tangential velocity components for the $i-$th star, respectively. Each component had its error, namely $\delta v_{i,{\rm R}}, \delta v_{i,{\rm T}}$. 
All the values were computed in mas/yr, independent of any assumption on the cluster distance.
Finally, the mean velocities $\langle v_{\rm R} \rangle$ and $ \langle v_{\rm T} \rangle$, the velocity dispersion components $\sigma_{\rm R}$ and $\sigma_{\rm T}$, and the 1D velocity dispersion ($\sigma_\mu$, see \secref{sec:results_simulation}) were computed through a Markov chain Monte Carlo (MCMC) exploration of the parameter space.
In particular, we used the \texttt{emcee} package\footnote{Full documentation available at  \url{https://emcee.readthedocs.io/en/stable/}} \citep{foreman-mackey_etal2013}, which provides a Python implementation of the affine-invariant MCMC sampler, enabling us to sample the posterior distribution.
The same analysis was carried out for stars within $0.2\rhl$ and around $\rhl$, and the 1D velocity dispersion ratio was computed.
For clusters with field of view (FoV) coverage smaller than $\rhl$, we opted for a hybrid approach. We estimated the inner 1D velocity dispersion using individual stars (see Eq.~\ref{eq:likelihood}) whereas we relied on dynamical modeling for the outer one.
Using a single-mass, King model \citep[][constructed via the LIMEPY\footnote{publicy available at \url{https://github.com/mgieles/limepy}} Python library developed by \citealt{gieles_zocchi2015}]{king_1966} we fitted both the density \citep{deBoer_etal2019} and the 1D velocity dispersion \citep[obtained by merging HST data from ][, and \emph{Gaia} DR3 data, \citealt{vasiliev_baumgardt2021}]{libralato_etal2022} profiles. The former allowed us to constrain the structural parameters (such as $W_0$, and $r_{\rm hm}$), while the latter constrained the total cluster mass. We discuss the fitting procedure and present the results in Appendix~\ref{appendix:additional_gcs}.

To quantify mass segregation, we defined the parameter $\deltaobsnew{\rlim}$: similarly to \myeqref{eq:delta_definition}, $\deltaobsnew{\rlim}$ quantifies the degree of segregation via the area between the nCRDs of a bright (BP) and faint (FP) population computed within a given distance, $\rlim$, from the center. 
We computed $\deltaobsnew{\rlim}$ using the photometric catalog provided by \citet{sarajedini_etal2007}. 

A critical step in this regard was a proper assessment of the photometric incompleteness of the catalogs. Due to crowding, incompleteness mainly affects faint stars, preferentially in the center. 
Given that $\deltaobsnew{\rlim}$ traces the mass segregation using the relative spatial distributions, incomplete catalogs bias the results in a non-trivial manner, possibly inflating the mass-segregation signal. 
We therefore estimated the completeness ($c$) for every star, accounting for its projected distance from the center and magnitude. 
Finally, each star contributed to the nCRD with a factor $1/c$.

\subsection{Accounting for incompleteness in evaluating $\deltaobsnew{\rlim}$\label{appendix:completeness}}
\begin{figure*}[!th]
    \centering
    \includegraphics[width=0.49\textwidth]{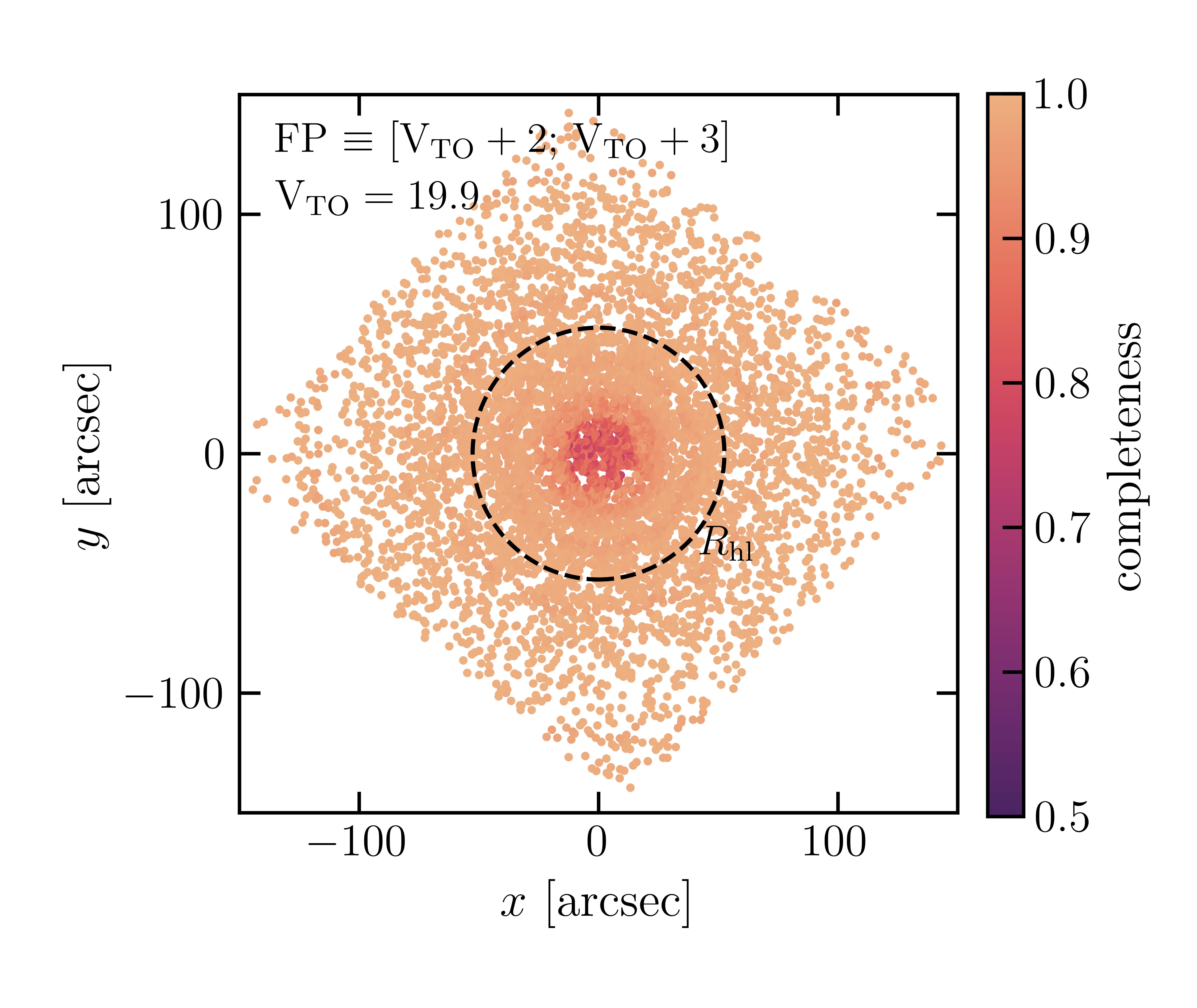}
    \includegraphics[width=0.49\textwidth]{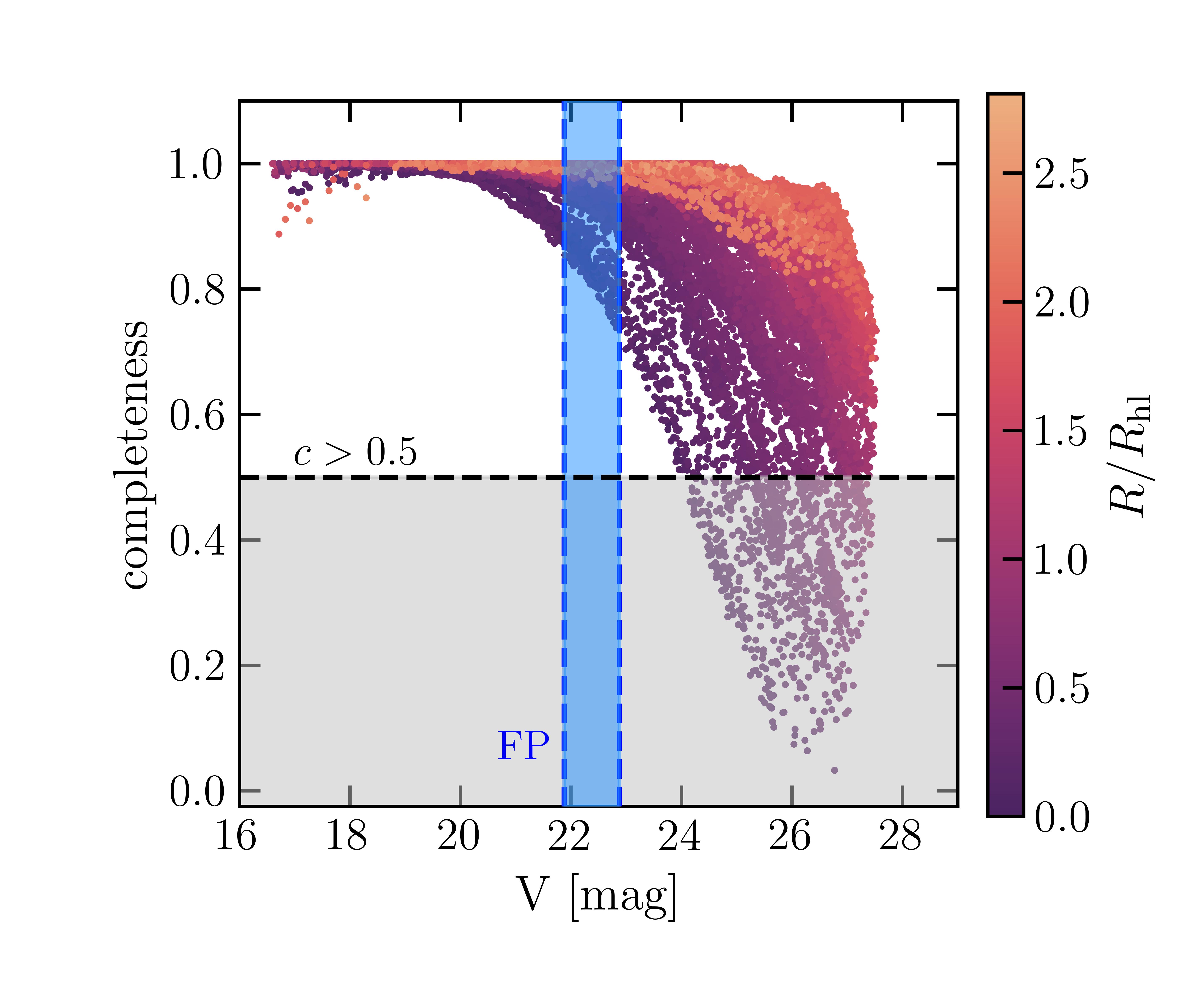}
    \caption{Photometric completeness for \ngc{6584}. Left panel: two-dimensional map of sources in the magnitude range $[ \vto+2; \vto+3 ]$, with $\vto$ being the turn-off magnitude. Each star is color-coded according to the completeness in the V band. Right panel: V-magnitude dependence of the photometric completeness. Stars are color-coded according to their projected distance from the center normalized to $\rhl$. The gray area marks $c<0.5$ whereas the blue region shows the $[ \vto+2; \vto+3 ]$ magnitude range.}
    \label{fig:completeness_ngc6584}
\end{figure*}
\begin{figure*}[!th]
    \centering
    \includegraphics[width=0.49\textwidth]{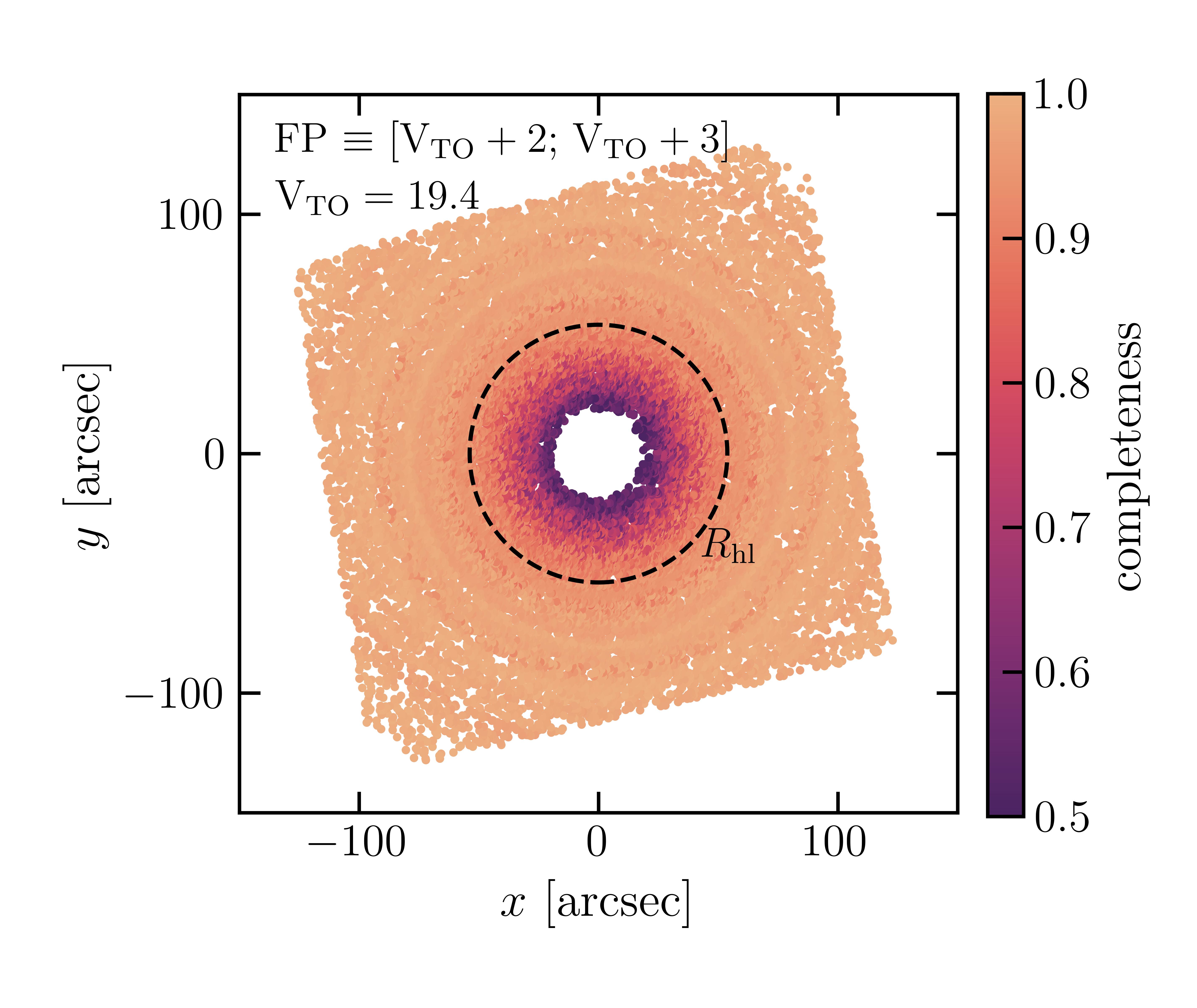}
    \includegraphics[width=0.49\textwidth]{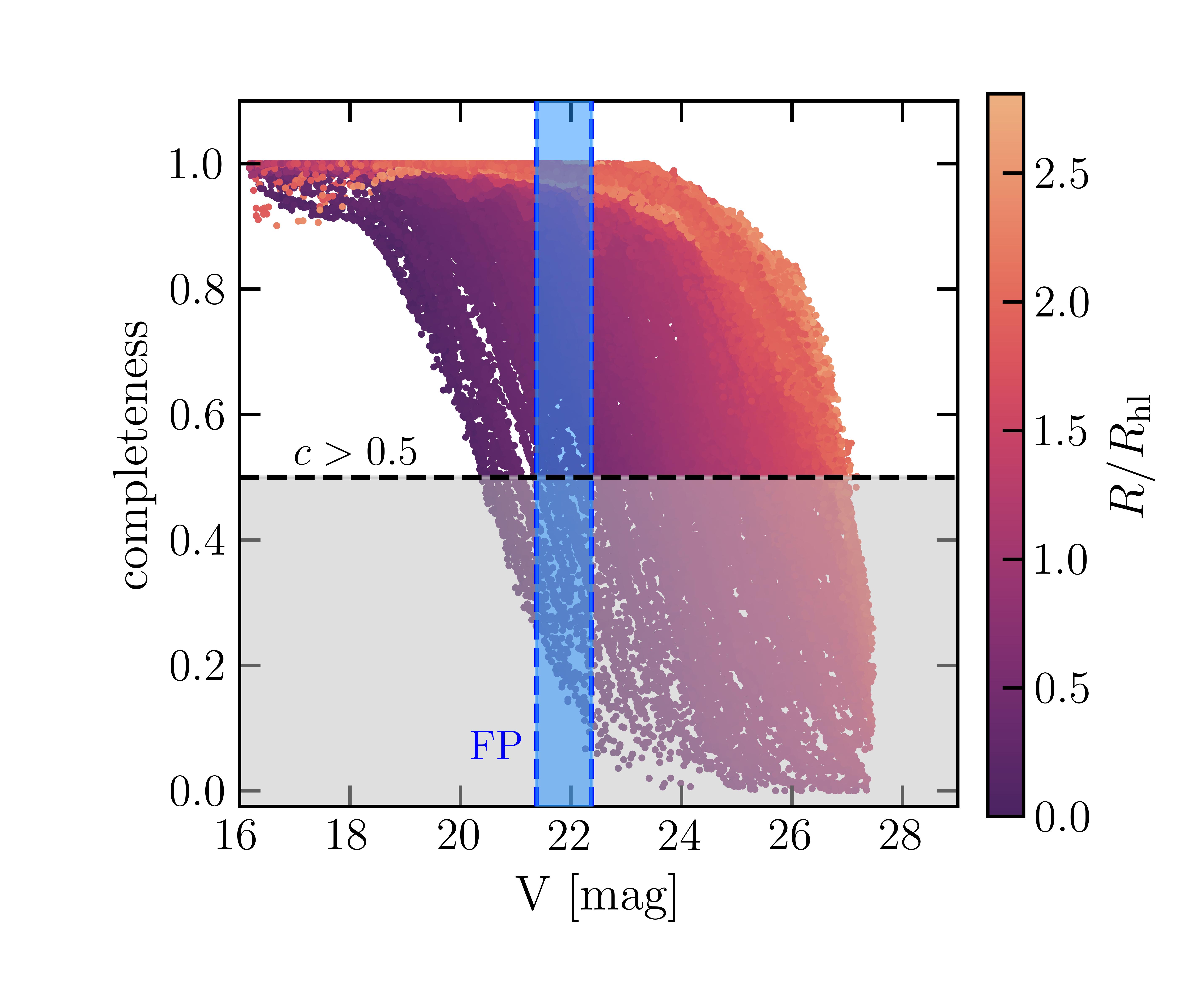}
    \caption{Same as \figref{fig:completeness_ngc6584} for \ngc{7089}.    }
    \label{fig:completeness_ngc7089}
\end{figure*}
In this Section, we present the details of the photometric completeness calculation, carried out for all the clusters in the sample.
In particular, we used the photometric catalog by \citet{sarajedini_etal2007}, and the artificial-star test catalog by \citet{anderson_etal2008} provided for each cluster.
The latter catalog consists of $10^5$ artificial stars, distributed uniformly within the core radius and with a density $\propto R^{-1}$ outside, with a flat luminosity function in the F606W filter and with colors along the cluster fiducial line \citep[see][ for further details]{anderson_etal2008}.

To estimate the completeness we adopted the following procedure:
\begin{enumerate}
    \item we determined quality selection criteria exploiting the \qfit parameter in both $V$ and $I$ bands as a function of magnitude. 
    Stars with a \qfit higher than the 90th percentile of the distribution at the star magnitude were not considered in the subsequent analyses;
    \item using stars with good photometry, we determined membership selection criteria in the color-magnitude diagram ($V$ versus $V-I$). Dividing the stars in magnitude bins (0.5 magnitudes wide), we found the 10th and 90th percentiles of the color distribution. Stars outside this range were not included, as probably field interlopers;
    \item we applied these selections to the observed and photometrically-calibrated artificial-star catalogs
    \footnote{For \ngc{6144} 
    we applied slightly different selections, adopting the 5th and 95th quantiles for the color-magnitude diagram and \texttt{QFIT} selections. Indeed, we found that the previous selections introduced systematics in the spatial distribution of the brightest stars thereby biasing the calculation of mass-segregation proxies through the nCRDs.}. 
    In this step, particular care was paid to keeping artificial stars without an output magnitude, which are those input stars not recovered by the data reduction pipeline;
    \item for each observed star we selected artificial stars within a radial shell centered around the star position. The shell width was iteratively widened until at least 1000 artificial stars were selected. 
    We then constructed the completeness curve as a function of the magnitude. 
    For each magnitude bin, the completeness was directly computed as the ratio between the number of recovered and input artificial stars. We considered a star as recovered if the output and input $V$ magnitudes were compatible with a tolerance of 0.75 magnitudes, due to photometric blends \citep[as suggested by ][]{anderson_etal2008}. The final completeness value assigned to each star was obtained by interpolating this curve and evaluating it at the star magnitude.
\end{enumerate}

We defined the BP as stars with V magnitude $[\vto;\,\vto+1]$, whereas FP stars have magnitudes in the range $[\vto+2;\,\vto+3]$. 
The limit of $V<\vto+3$ ensured that $c>0.5$ over the whole FoV while keeping the mass gap between BP and FP stars as large as possible (see also discussion in \secref{sec:galactic_gc_properties}).
Lower completeness estimates are indeed more uncertain as the completeness relative error roughly scales as the inverse of the square root of the number of recovered stars. Therefore, retaining stars with very low completeness makes the nCRD more uncertain.
Also, evaluating the impact of pushing to very low completeness regimes is not straightforward, as one might over- or under-estimate the nCRD due to significant statistical fluctuations in the completeness estimate.

In \figref{fig:completeness_ngc6584} we show the completeness properties of FP stars in \ngc{6584}: 
as expected, the $c$ decreases toward the center and for fainter magnitudes. However, the FP stars in this cluster are characterized by completeness levels $\gtrsim 75\%$ at all distances.

In \figref{fig:completeness_ngc7089} we show the 2D radial distribution (left panel) and completeness variation curves (right panel) for \ngc{7089} as a prototypical case of a cluster excluded from our analysis. While this GC fulfills the radial coverage requirements (\secref{sec:galactic_gc_properties}), its FP is characterized by a strong incompleteness in the innermost regions. In fact, for this sub-sample of stars $c$ decreases rapidly towards the center, dropping well below the critical threshold of $0.5$ around $\rhl$, and almost no stars are found within $<0.5\rhl$.    
Such severe incompleteness makes the calculation of $\deltaobsnew{\rlim}$ practically unfeasible.
We note that a few similar cases (for example \ngc{2808}, and \ngc{6093}) were included in the analysis by \citet{weatherford_etal2020} with a possible significant impact on the derived values of $\Delta$ for these systems. 


\subsection{Comparison with simulations\label{sec:comparison_obs_sim}}
\begin{figure*}[!th]
    \centering
    \includegraphics[width=0.47\textwidth]{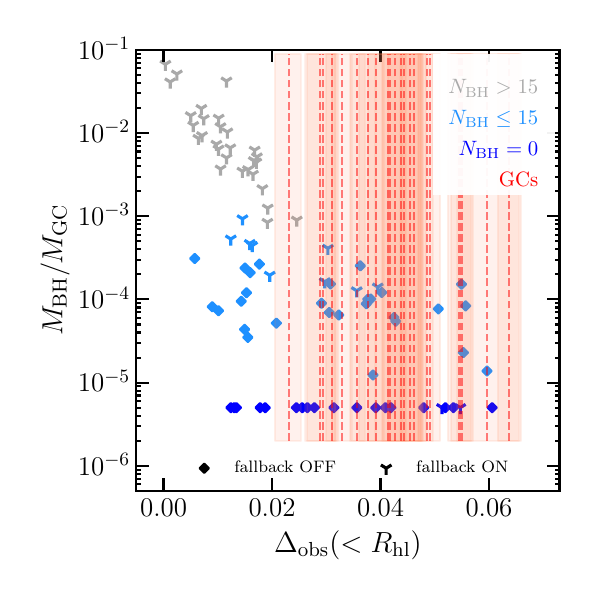}
    \includegraphics[width=0.47\textwidth]{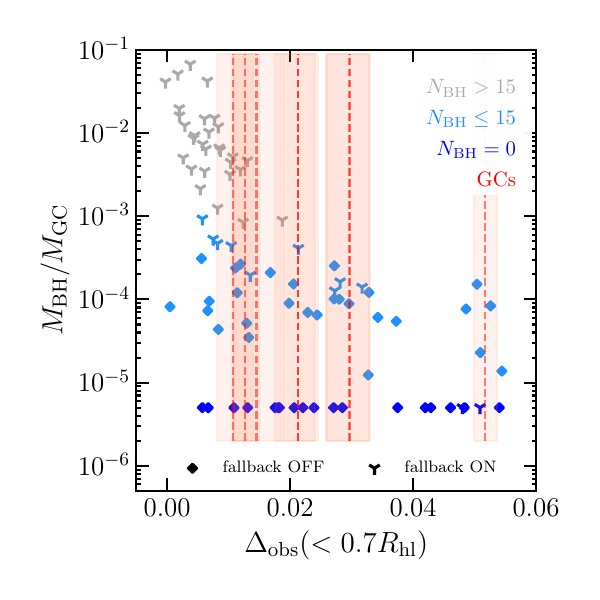}
    \caption{BH mass fraction as a function of $\deltaobs$ (left panel) and $\deltaobsadditional$ (right panel). Points show the simulation properties 
    recomputed according to the magnitude and spatial selections adopted for the observations
    (see \secref{sec:galactic_gc_properties}), whereas the vertical red lines show values obtained for the Galactic GCs studied in this work (along with errors as shaded areas). Simulations without BHs at \gyr are shown at $\mbh/\mgc = 5\times10^{-6}$.}
    \label{fig:bhmass_vs_delta_observations_21cl}
\end{figure*}

\begin{figure*}[!th]
    \centering
    \includegraphics[width=0.47\textwidth]{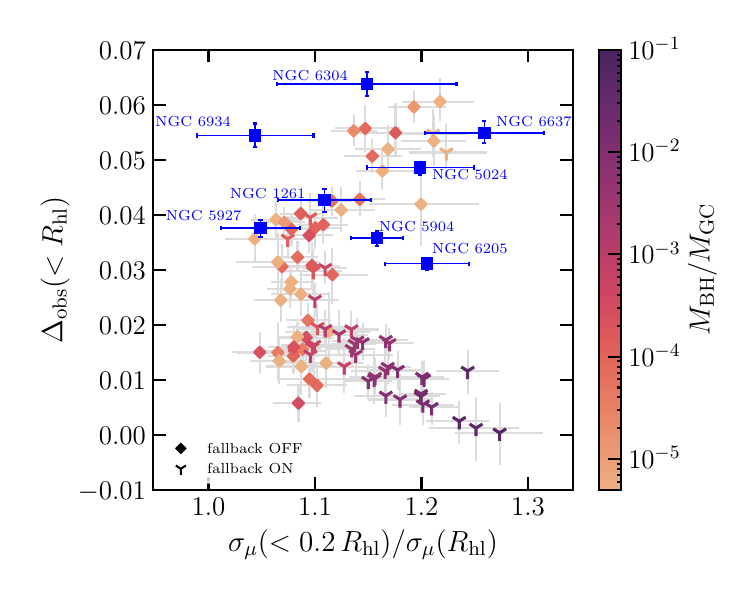}
    \includegraphics[width=0.47\textwidth]{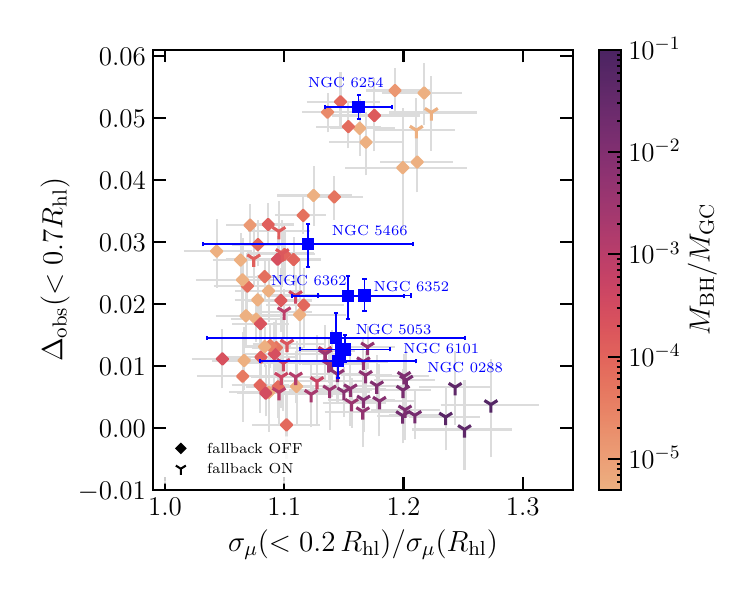}
    \caption{$\deltaobs$ (left panel) and $\deltaobsadditional$ (right panel) as a function of the velocity dispersion ratio.
    Values from the simulations were recomputed according to the magnitude and spatial selections adopted for the observations (see \secref{sec:galactic_gc_properties}).
    Simulations are color-coded according to the BH mass fraction (see the color bar). Error bars were computed from multiple LOS projections. Finally, in blue we show the values (along with error bars) obtained for Galactic GCs (see \secref{sec:observations} for details). The observations shown in this plot are reported in \tabref{tab:results_gcs} and \tabref{tab:results_additional_gcs}.}
    \label{fig:delta_vs_veldispration_observationsN100}
\end{figure*}



Here, we compare the results obtained from the simulations (\secref{sec:results_simulation}) and the state-of-the-art data presented in \secref{sec:galactic_gc_properties}.

\figref{fig:bhmass_vs_delta_observations_21cl} shows the BH mass fraction as a function of $\deltaobsnew{\rlim}$ for $\rlim=\rhl$ (left panel) and $\rlim=0.7\rhl$ (right panel). Values from both simulations and observations were computed according to the definition in \secref{sec:galactic_gc_properties}, adopting $\rlim = \rhl$ ($0.7\rhl$ for those GCs with smaller FoV coverage), and using BP and FP stars to compute $\Delta$.
The simulations cover a similar range of $\deltaobsnew{\rlim}$ as the observations. In addition, each value of $\deltaobsnew{\rlim}$ could be reproduced by either simulation with high or low BH mass fractions. This further highlights the degeneracies and strengthens the need for a multi-dimensional approach.

\figref{fig:delta_vs_veldispration_observationsN100} shows $\deltaobsnew{\rlim}$-vs-$\vdispratio$. 
Simulations are color-coded according to the BH mass fraction, with different symbols depicting whether the fallback prescription was adopted for the BH natal kicks.
Finally, blue points show the values obtained for Galactic GCs (see \secref{sec:galactic_gc_properties}). 
In particular, we show only those clusters for which at least 100 stars with kinematics were available in the radial ranges considered. Such a threshold ensures that the errors on $\vdispratio$ are small enough for a meaningful comparison with the simulations. 
Nonetheless, in \tabref{tab:results_gcs} we provide the values of $\deltaobsnew{\rlim}$ for all the clusters, and $\vdispratio$ (and relative errors) for the clusters shown in \figref{fig:delta_vs_veldispration_observationsN100}.

The left panels of Figs.~\ref{fig:bhmass_vs_delta_observations_21cl}, and \ref{fig:delta_vs_veldispration_observationsN100} show GCs for which the data coverage was $\geq\rhl$.
For these clusters
the $\deltaobs$ and $\vdispratio$ values are in reasonable agreement with numerical-simulation predictions except for a few cases that show discrepant values of $\deltaobs$. Also, they typically show $\deltaobs\gtrsim 0.03$ (see also \tabref{tab:results_gcs}). 
The mass segregation and the kinematic properties of these clusters could be thus reproduced either by systems in which BHs were ejected right after formation (due to high natal kicks) or that lose their BHs due to dynamical interactions in the center. In either case, the present-day BH mass fraction is likely low \citep[similarly to what was found by][]{weatherford_etal2020,dickson_etal2024}.

As discussed in \secref{sec:galactic_gc_properties}, we also considered GCs with a FoV coverage smaller than $\rhl$.
Within this sample, a few notable clusters were indeed suggested to host massive BH populations at their center (e.g., \ngc{5053}, \ngc{6101}, and \ngc{6362}, see \citealt{askar_etal2018,weatherford_etal2020}).
The right panel of \figref{fig:bhmass_vs_delta_observations_21cl} shows the $\deltaobsadditional$ values obtained for these GCs. 
As expected from previous works \citep{dalessandro_etal2015,peuten_etal2016,weatherford_etal2020} 
some of these clusters exhibit little mass segregation. This feature could be interpreted as either the result of the BH burning phase \citep{kremer_etal2020} or slow dynamical evolution, as already discussed in \secref{sec:results_simulation}. 

In the right panel of \figref{fig:delta_vs_veldispration_observationsN100}, we delve more into the kinematic properties of these clusters using $\vdispratio$ introduced in this work:
focussing on clusters with little mass segregation (roughly $\deltaobsadditional<0.02$, see the right panel of \figref{fig:delta_vs_veldispration_observationsN100}) we notice that while there might be hints of higher values of $\vdispratio$ (which would imply these GCs host a nonnegligible BH mass fraction), observational errors do not allow us to discriminate between the possible scenarios fully. 

Finally, we highlight here that decreasing the radial and mass ranges for the calculation of $\Delta$-like quantities almost hampers a proper distinction between systems with or without BHs using the $\vdispratio$ ratio (see Figs.~\ref{fig:delta_veldispratio}, and \ref{fig:delta_vs_veldispration_observationsN100}). Hence, future surveys covering larger radial and mass ranges would be critical in this respect.

\section{Conclusions\label{sec:conclusion}}
In this work, we tackled the inference of the BH mass fraction in GCs through observable properties. We used a survey of Monte Carlo simulations exploring a large range of initial conditions and different prescriptions for the BH natal kicks.
We demonstrated that single observables such as parameters measuring the degree of mass segregation are not suited for inferring
the BH mass fraction in real GCs, because of significant degeneracies.
This degeneracy naturally arises because clusters without a sizable BH population but being dynamically younger may exhibit similar features (e.g., in terms of mass segregation features) when compared to systems where the dynamical evolution was halted by the BH burning mechanism.
This highlights that the role of possible different initial conditions and physical assumptions should be carefully considered when trying to obtain the present-day BH population in Galactic GCs.

We then explored multiple probes that could help us break this degeneracy.  
We introduced the combination $\Delta$ and $\vdispratio$ as a possible candidate pair. 
$\Delta$ traces the mass segregation of visible stars within the clusters, whereas $\vdispratio$ quantifies the steepness of the velocity dispersion profile: the presence of a massive BH subsystem increases $\vdispratio$ while halting the mass segregation of visible stars (i.e., keeping $\Delta$ low). At the same time, dynamically young clusters (i.e., exhibiting a low degree of mass segregation) that did not retain a massive BH population at \gyr, have typically lower $\vdispratio$.

We therefore measured $\deltaobsnew{\rlim}$ (assuming either $\rlim = \rhl$ or $\rlim = 0.7\rhl$, see \secref{sec:comparison_obs_sim}) and $\vdispratio$ for several Galactic GCs using the photometric and astrometric catalogs by \citet{sarajedini_etal2007} and \citet{libralato_etal2022} respectively, and we compared them with the same quantities computed from the simulations. We found that current state-of-the-art data do not provide stringent enough constraints to fully discriminate between different scenarios, likely due to the limited radial and mass ranges.
Future astrometric and photometric data provided by, for instance, the \emph{Roman} space telescope \citep{roman_whitepaper} may allow us to shed light on the subject.

Finally, we also presented a detailed discussion on the calculation of the photometric completeness using artificial star tests. We found that for a non-negligible number of clusters, the calculation of $\deltaobsnew{\rlim}$ was not feasible due to severe incompleteness in the center. Some of these clusters were previously studied using the same photometric catalog 
to infer the BH mass fraction \citep[see e.g.,][]{weatherford_etal2020}.
We thus advise caution in interpreting those results.

In summary, 
we showed that the effects of BHs on the internal GC dynamics over their lifetime cannot be encapsulated in a single observable, thus multiple physical properties should be used to infer the present-day BH populations in real GCs.

\begin{acknowledgements}
The authors thank Ata Sarajedini, and Mattia Libralato for sharing some private data used in this work.
A.D.C. and E.D. are also grateful for the warm hospitality of Indiana University where part of this work was performed.
E.D. acknowledges financial support from the Fulbright Visiting Scholar Program 2023.
E.V. acknowledges support from the John and A-Lan Reynolds Faculty Research Fund.
This research was supported in part by Lilly Endowment, Inc., through its support for the Indiana University Pervasive Technology Institute.
\end{acknowledgements}

\bibliographystyle{aa}
\bibliography{bibfile}

\appendix

\section{Density distribution and velocity dispersion profiles for nine GCs \label{appendix:additional_gcs}}
\begin{figure*}[!th]
    \centering
    \includegraphics[width=\textwidth]{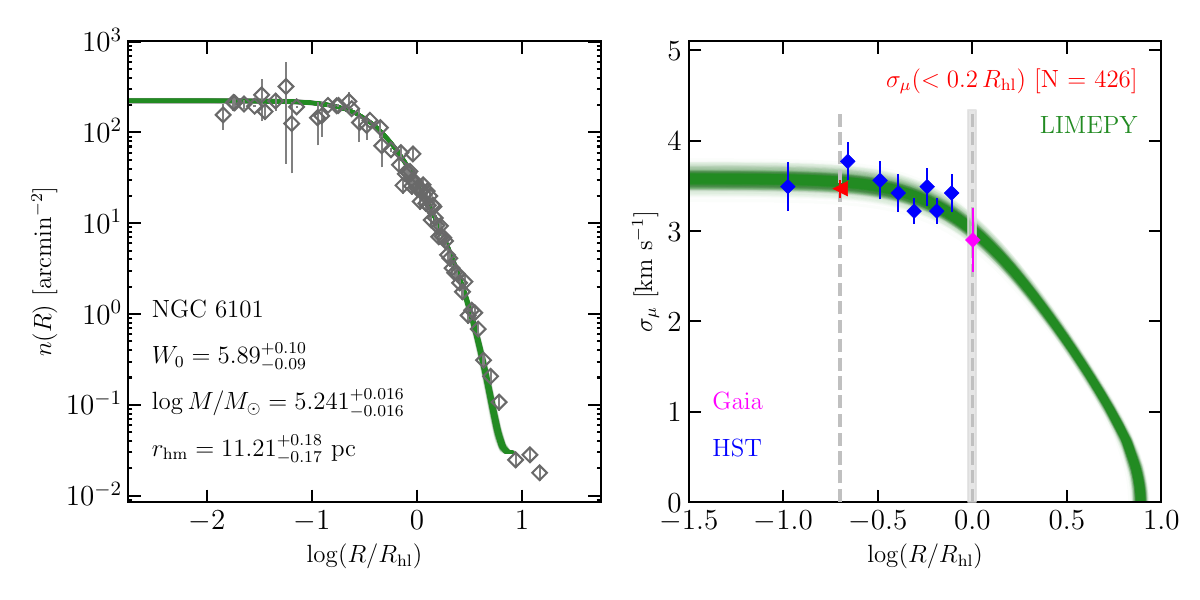}
    \caption{Structural and kinematical properties for \ngc{6101}. Left panel: number density profile from \citet[][in gray]{deBoer_etal2019}. Posterior values for the model's free parameters are reported. The green lines show 1000 model realizations from posterior samples. Right panel: 1D velocity dispersion profile in \kms. In blue are HST data from \citet{libralato_etal2022}, whereas in purple are \emph{Gaia} data from \citet{vasiliev_baumgardt2021}. Velocities were converted assuming the distance from \citet{baumgardt_vasiliev2021}. The red point shows the $\sigma_\mu$ obtained from single stars $<0.2\rhl$ plotted at $0.2\rhl$ (the number of stars used in the calculation is also reported within square brackets). The vertical lines mark $0.2\rhl$, and $\rhl$ with the light-gray band depicting the radial range $[0.95;\,1.05]~\rhl$.}
    \label{fig:limepy_fit_ngc6101}
\end{figure*}
In this Appendix, we present
the hybrid approach in the $\vdispratio$ calculation adopted for a subsample of GCs (see \secref{sec:observations}).
We used the number density profiles provided by \citet{deBoer_etal2019}. The authors stiched heterogeneous profiles from literature, such as surface brightness \citep{trager_eral1995}, and number density \citep{miocchi_etal2013} profiles, complemented by \emph{Gaia} data.
For the 1D velocity dispersion profiles, we used the catalogs by \citet{libralato_etal2022} and \citet{vasiliev_baumgardt2021} to ensure a larger radial coverage. 

We fitted these profiles with a single-mass King model constructed using the LIMEPY \citep{gieles_zocchi2015} Python library. 
Within a Bayesian framework, we assumed the likelihood function
\begin{equation}
    \ln \mathcal{L} = \ln \mathcal{L}_{\rm profile} + \ln \mathcal{L}_{\rm vel. disp.} \,,
    \label{eq:likelihood_limepy}
\end{equation}
where the first and second terms of the right-hand side of \myeqref{eq:likelihood_limepy} are the likelihoods for the density and the velocity dispersion profiles, respectively.
Concerning the number density, the likelihood term is
\begin{equation}
    \ln \mathcal{L}_{\rm profile} = -\frac{1}{2} \sum^{N_p}_{i=1} \frac{(n_i - \eta \Sigma(R_i|\boldsymbol{\theta}))^2}{\delta n^2_i} \,,
\end{equation}
with $R_i$, $n_i$, and $\delta n_i$ being the projected cluster-centric distance, the number density, and the relative error for all $N_p$ bins. The projected mass density ($\Sigma$) was computed at $R_i$ for any given set of the model's free parameters, $\boldsymbol{\theta} = \{ W_0, \log M_{\rm GC}, r_{r\rm hm} \}$, namely the dimensionless central potential, cluster mass, and half-mass radius \citep{king_1966}. Finally, $\eta$ is a nuisance parameter for scaling the mass-density profile into the number-density one.
The velocity dispersion ($\sigma_\mu$) term in \myeqref{eq:likelihood_limepy}
\begin{equation}
    \ln \mathcal{L}_{\rm vel.disp.} = -\frac{1}{2} \sum^{N_\sigma}_{i=1} \frac{(\sigma_{\mu\,,i} - \sigma_{\mu\,,{\rm model}}(R_i|\boldsymbol{\theta}))^2}{\sigma_{\mu\,,i}^2} \,,
\end{equation}
summed over the $N_\sigma$ bins. The velocity dispersion from the model ($\sigma_{\mu\,,{\rm model}}$) was computed at any given radial position $R_i$ for each set of the model's free parameters.

We explore the free parameters space using an MCMC approach exploiting the Python implementation provided by the \texttt{emcee} \citep{foreman-mackey_etal2013} library. For each cluster, we used 100 walkers, evolved for 500 steps. The first quarter was discarded for the sake of convergence and one sample every 50 was retained to account for correlations.
In \figref{fig:limepy_fit_ngc6101} we show the number density and velocity dispersion profiles for \ngc{6101}. Posterior values for the model's free parameters, as well as 1000 models constructed from posterior samples, are also shown.

Within the hybrid approach, $\sigma_\mu(<0.2\rhl)$ is computed using single stars (as presented in \secref{sec:observations}, see Eq.~\ref{eq:likelihood}), whereas $\sigma_\mu(\rhl)$ (and its relative error from the 16th and 84th percentiles of the posterior distribution on the velocity dispersion profile) is computed from the dynamical modeling at $R=\rhl$. In doing so, we also verified that the value does not significantly change (within the errors) if computed at $R=0.95\rhl$ or $R=1.05\rhl$, which are the boundaries of the radial shell used in the single star analysis (see \secref{sec:observations}). 
Finally, the 1D velocity dispersion ratio is computed.

\FloatBarrier
\section{Table of values of $\Delta$s and velocity dispersion ratios \label{appendix:table_results}}
\renewcommand{\arraystretch}{1.5}
\begin{table}[!th]
    \centering
    \caption{Properties of the sample of 21 clusters analyzed in this study with FoV larger than $\rhl$. \label{tab:results_gcs}}
    \begin{tabular}{lcc}
    \hline
    Cluster & $\deltaobs$ & $\vdispratio$ \\
    \hline \hline
    \ngc{1261}& $0.043^{+0.002 }_{-0.002 }$ & $1.11 \pm 0.04$ \\
    \ngc{2298}& $0.044^{+0.004 }_{-0.003 }$ & $-$ \\
    \ngc{4590}& $0.049^{+0.002 }_{-0.002 }$ & $-$ \\
    \ngc{4833}& $0.046^{+0.002 }_{-0.002 }$ & $-$ \\
    \ngc{5024}& $0.049^{+0.001 }_{-0.001 }$ & $1.20 \pm 0.05$ \\
    \ngc{5904}& $0.036^{+0.001 }_{-0.001 }$ & $1.16 \pm 0.02$ \\
    \ngc{5927}& $0.038^{+0.001 }_{-0.002 }$ & $1.05 \pm 0.04$ \\
    \ngc{5986}& $0.039^{+0.001 }_{-0.002 }$ & $-$ \\
    \ngc{6144}& $0.029^{+0.003 }_{-0.003 }$ & $-$ \\
    \ngc{6171}& $0.044^{+0.003 }_{-0.003 }$ & $-$ \\
    \ngc{6205}& $0.031^{+0.001 }_{-0.001 }$ & $1.21 \pm 0.04$ \\
    \ngc{6218}& $0.046^{+0.002 }_{-0.002 }$ & $-$ \\
    \ngc{6304}& $0.064^{+0.002 }_{-0.002 }$ & $1.15 \pm 0.08$ \\
    \ngc{6535}& $0.042^{+0.006 }_{-0.007 }$ & $-$ \\
    \ngc{6584}& $0.023^{+0.002 }_{-0.003 }$ & $-$ \\
    \ngc{6637}& $0.055^{+0.002 }_{-0.002 }$ & $1.26 \pm 0.06$ \\
    \ngc{6717}& $0.060^{+0.006 }_{-0.006 }$ & $-$ \\
    \ngc{6723}& $0.033^{+0.002 }_{-0.002 }$ & $-$ \\
    \ngc{6779}& $0.041^{+0.002 }_{-0.002 }$ & $-$ \\
    \ngc{6934}& $0.054^{+0.002 }_{-0.002 }$ & $1.04 \pm 0.05$ \\
    \ngc{6981}& $0.029^{+0.002 }_{-0.003 }$ & $-$\\
    \hline
    \end{tabular}
    \begin{tablenotes}
    \item[]\textbf{Notes.} $\deltaobs$ and velocity dispersion ratio for the clusters with FoV coverage of at least one $R_{\rm hl}$. The velocity dispersion was computed directly from individual stars within $0.2 \rhl$ and around $\rhl$. Only clusters with at least 100 stars within these ranges were considered.
    \end{tablenotes}
\end{table}

\begin{table}[!th]
    \centering
    \caption{Properties of nine clusters with FoV coverage smaller than $\rhl$. \label{tab:results_additional_gcs}}
    \begin{tabular}{lcc}
    \hline
    Cluster & $\deltaobsadditional$ & $\vdispratio$ \\
    \hline \hline
    \ngc{0288}& $0.011^{+0.003 }_{-0.003 }$ & $1.15 \pm 0.07$ \\
    \ngc{6254}& $0.052^{+0.002 }_{-0.002 }$ & $1.16 \pm 0.03$ \\
    \ngc{6352}& $0.021^{+0.003 }_{-0.004 }$ & $1.15 \pm 0.05$ \\
    \ngc{6362}& $0.021^{+0.003 }_{-0.002 }$ & $1.17 \pm 0.04$ \\
    \ngc{6496}& $0.030^{+0.003 }_{-0.004 }$ & $-$ \\                
    \ngc{6752}& $0.076^{+0.002 }_{-0.002 }$ & $1.35 \pm 0.02$ \\
    \ngc{5053}& $0.015^{+0.004 }_{-0.004 }$ & $1.14 \pm 0.11$ \\
    \ngc{5466}& $0.030^{+0.003 }_{-0.004 }$ & $1.12 \pm 0.09$ \\
    \ngc{6101}& $0.013^{+0.002 }_{-0.002 }$ & $1.15 \pm 0.04$ \\
    \hline
    \end{tabular}
    \begin{tablenotes}
    \item[]\textbf{Notes.} $\deltaobsadditional$ and velocity dispersion ratio for the clusters with FoV coverage $< R_{\rm hl}$. The velocity dispersion ratio was computed using individual stars (within $0.2 \rhl$) complemented by dynamical modeling around $\rhl$. For \ngc{6496} the $\vdispratio$ value was not computed due to large errors in the PMs that hindered a reliable calculation of $\sigma_\mu$.
    \end{tablenotes}
\end{table}
In this Appendix, we report the results for the GC sample considered in this study.
In particular, in \tabref{tab:results_gcs} we list the values $\deltaobs$ and $\vdispratio$ obtained for the sample of 21 clusters with HST coverage of at least one half-light radius (see \secref{sec:observations} for the definition and calculation details of the parameters).
Similarly, \tabref{tab:results_additional_gcs} shows the results (namely $\deltaobsadditional$, and $\vdispratio$) for the nine clusters with HST coverage smaller than one half-light radius.

\end{document}